\documentclass{article}

 \usepackage[preprint]{neurips_2025}
\usepackage{booktabs}
\usepackage{graphicx}   
\usepackage{booktabs}  
\usepackage[utf8]{inputenc} 
\usepackage[T1]{fontenc}    
\usepackage{hyperref}       
\usepackage{url}            
\usepackage{booktabs}       
\usepackage{amsfonts}       
\usepackage{nicefrac}       
\usepackage{microtype}      
\usepackage{xcolor}         
\usepackage{float} 
\usepackage{multirow}
\usepackage{caption}
\usepackage{amsmath}
\title{\LARGE Rainbow Noise: Stress-Testing Multimodal Harmful-Meme Detectors on LGBTQ Content}

\author{
  Ran Tong \\
  Mathematics and Statistics Department \\
  University of Texas at Dallas \\
  \texttt{rxt200012@utdallas.edu} \\
  \And
  Songtao Wei \\
  Computer Science Department \\
  University of Texas at Dallas \\
  Richardson, TX, USA \\
  \texttt{songtao.wei@utdallas.edu} \\
  \And
  Jiaqi Liu \\
  Independent Researcher \\
  \texttt{jackyliu9747@gmail.com} \\
  \And
  Lanruo Wang \\
  Naveen Jindal School of Management \\
  University of Texas at Dallas \\
  Richardson, TX 75080 \\
}

\begin{document}

\maketitle

\begin{abstract}
Hateful memes aimed at LGBTQ\,+ communities often evade detection by
tweaking either the caption, the image, or both.  
We build the first robustness benchmark for this setting, pairing four
realistic caption attacks with three canonical image corruptions and
testing all combinations on the PrideMM dataset.  
Two state-of-the-art detectors, MemeCLIP and MemeBLIP2, serve
as case studies, and we introduce a lightweight \textbf{Text Denoising Adapter (TDA)} to enhance the latter's resilience.
Across the grid, MemeCLIP degrades more gently, while MemeBLIP2 is particularly sensitive to the caption edits that disrupt its language processing. 
However, the addition of the TDA not only remedies this weakness but makes MemeBLIP2 the most robust model overall.
Ablations reveal that all systems lean heavily on text, but architectural choices and pre-training data significantly impact robustness.
Our benchmark exposes where current multimodal safety models crack and demonstrates that targeted, lightweight modules like the TDA offer a powerful path towards stronger defences.
\end{abstract}

\section{Introduction}
\noindent

Memes—images with brief captions—spread rapidly online and shape public conversation. While many are humorous or activist, some carry hate or harassment that fuels prejudice and threatens online safety. Detecting these harmful memes is therefore a critical research goal.~\cite{liu2025memeblip2}.

\indent
\noindent
This need is especially pressing for LGBTQ users, who face higher levels of online abuse than the general population. A Pew Research Center survey reported that \(69\%\) of lesbian, gay, and bisexual adults who had used online-dating platforms encountered at least one form of harassment, compared with \(52\%\) of their straight counterparts~\cite{pew2020harassment}. A 2022 safety report led by GLAAD found that \(60\%\) of LGBTQ respondents felt harmed not only by direct abuse but also by witnessing hate aimed at other community members~\cite{glaad2022safety}. 

Attackers often mask hate by scribbling on the picture, covering key regions or swapping letters for look-alike symbols. These edits confuse multimodal models, so we must measure how well harmful-meme detectors handle realistic text and image noise to keep LGBTQ users safe. 

In this paper we mainly evaluate two recent light-weight multimodal detectors— 
 MemeCLIP~\cite{shah2024memeclip} and MemeBLIP2~\cite{liu2025memeblip2}—both of which learn joint image–text embeddings and are trained on the PrideMM corpus for the specific task of spotting hateful or harassing memes directed at LGBTQ+ communities. we also benchmark them against GPT-4.1 Vision, a state-of-the-art, general-purpose Large Multimodal Model (LMM). This comparison allows us to assess whether efficient, task-specific models can achieve results comparable to one of the most advanced systems available.


Shah et al. introduced MemeCLIP, a model designed for efficient meme classification \cite{shah2024memeclip}. It is built on the powerful CLIP model \cite{radford2021learning}, which is  excellent at understanding how images and text relate. MemeCLIP keeps the main CLIP model frozen to preserve its knowledge and adds small, lightweight modules, including "Feature Adapters," that are trained specifically to detect harmful memes. This approach allows MemeCLIP to effectively leverage CLIP's strong, pre-existing alignment between vision and language.
Later, Liu et al. proposed MemeBLIP2, which uses the more advanced BLIP-2 backbone \cite{li2023blip2}. Unlike CLIP, the image and text encoders in BLIP-2 are not inherently aligned. MemeBLIP2 addresses this by using a similar lightweight strategy, adding trainable projection layers and adapter modules to connect the frozen encoders. The goal of MemeBLIP2 is to capture more subtle and ironic cues that are common in memes.
Together, these works show a new direction for AI safety, shifting the focus from large, general models to lightweight systems that efficiently adapt pre-trained knowledge. This strategy is highly effective, as both MemeCLIP and MemeBLIP2 outperform general-purpose baselines like BERT and GPT-4o in Accuracy, AUROC, and F1 score on the PrideMM dataset \cite{shah2024memeclip}.

PrideMM comprises 5,063 memes related to the LGBTQ+ movement, each consisting of an image overlaid with text. Every meme is annotated for (i) hate-speech presence, (ii) target group (for example, gay men, trans women, or the broader LGBTQ community), (iii) stance (supportive, neutral, hostile), and (iv) the presence of humor. The images were collected from social platforms between 2017 and 2023 and were manually de-duplicated and split into train/validation/test partitions that keep label distributions balanced. 

Although both MemeCLIP and MemeBLIP2 perform well on clean data, recent studies show that vision--language models can lose substantial accuracy when either the caption or the image is perturbed \cite{goyal2023robustness}.  
This observation leads us to three research questions: 

(i) How much do key metrics drop under realistic caption or image noise?

(ii) Which channel is chiefly responsible for the loss?

(iii) Which perturbation families, alone or combined, are most damaging?

(iv)What architectural interventions can be designed to restore the models' robustness to these perturbations?

Answering these questions will reveal the limits of current multimodal detectors and indicate where robustness efforts should concentrate. Our main contributions are: 

First, we construct the first comprehensive robustness benchmark for the PrideMM dataset by evaluating multiple state-of-the-art models across a full grid of combined text and image perturbations. This benchmark not only allows for a direct comparison of model resilience but also pinpoints the specific combinations of noise that cause the most significant performance degradation.

Second, by isolating the impact of noise on each modality, we show that both MemeCLIP and MemeBLIP2 depend far more on the caption than the image. Our experiments reveal that corrupting the text is significantly more damaging to model performance than corrupting the visual content.

Third, to address the robustness gap we identified in MemeBLIP2,we integrate a lightweight Text Denoising Adapter (TDA) into the model's architecture. Our experiments demonstrate that this module significantly enhances MemeBLIP2's robustness, establishing the TDA as an effective and targeted tool for improving the resilience of such multimodal models.

To better help the LGBTQ+ community, we are releasing all code, perturbation scripts, and pre-trained checkpoints as well as PrideMM-Aug, a new dataset of 10,000 samples generated through a combination of realistic text and image augmentation strategies. This dataset is designed to serve as a resource for robustness fine-tuning, enabling researchers to harden their models against perturbations.
\section{Related Work}
Early multimodal safety studies focused on single–channel benchmarks such as
the Facebook \emph{Hateful Memes} challenge, which mixes images with overlaid
text and forces a classifier to process both to succeed
\cite{kiela2020hateful}.  
Subsequent datasets broadened the topic: Memotion~\cite{sharma2020memotion}
adds sentiment labels, MMHS150K~\cite{zhu2020m3h} scales to 150\,000 Twitter
memes, and Polyjuice-GIR~\cite{hayes2021multimodalgir} pairs each hateful meme
with a benign twin so that shallow surface cues no longer resolve the task.Most benchmarks, however, assume clean input.  
Work on robustness is more fragmentary.  

Initial studies on model robustness often focused on the vision channel. Hendrycks \textit{et al.} introduced ImageNet-C and ImageNet-A to standardise common corruptions and natural adversarial examples for pure vision models~\cite{hendrycks2019benchmarking}. Kim \textit{et al.} later adapted these ideas to vision–language pre-training by adding five synthetic distortions and measuring the drop on VQA and retrieval tasks~\cite{kim2022vlcorrupt}.

Parallel research has explored the impact of noise in the text channel. Belinkov and Bisk showed that spelling errors alone can erase most learned knowledge in neural machine translation \cite{belinkov2018synthetic}; later studies such as HotFlip~\cite{ebrahimi2018hotflip} and Universal Triggers~\cite{wallace2019universal} demonstrated targeted caption attacks that transfer to CLIP-based models.

Comprehensive multimodal robustness suites are only now emerging, with benchmarks like MM-Robustness~\cite{qiu2024mmrobustness} and RMT~\cite{schiappa2022robustness} testing models on generic tasks. Other recent work explores robustness from a different angle, introducing tasks like Defeasible Visual Entailment to evaluate how models revise their interpretations when presented with new information that can strengthen or weaken an initial premise~\cite{zhang2025defeasible}. Other recent work explores robustness from a different angle, introducing tasks like Defeasible Visual Entailment to evaluate how models revise their interpretations when presented with new information that can strengthen or weaken an initial premise~\cite{zhang2025defeasible}. Beyond safety benchmarks, robustness and generalisation have also been investigated in federated learning on non-IID data, long-tail multi-label classification, and open-world detection~\cite{xiao2024confusion,xiao2025curiosity,wang2025c3owd}. Our study differs by focusing specifically on the \emph{hate-meme} domain and LGBTQ safety, and by creating a full combinatorial grid of perturbations to isolate interaction effects that earlier work did not.Our study differs by focusing specifically on the \emph{hate-meme} domain and LGBTQ safety, and by creating a full combinatorial grid of perturbations to isolate interaction effects that earlier work did not.


\section{Methodology}

\subsection{Image Noise Generation}
\begin{figure}[h]
\centering
\setlength{\tabcolsep}{4pt} 
\begin{tabular}{cc}
\includegraphics[width=0.46\linewidth]{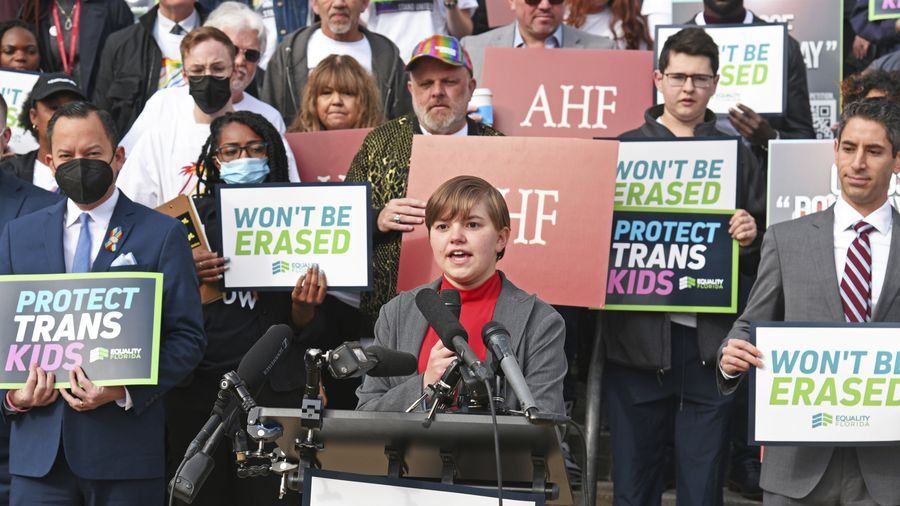 } &
\includegraphics[width=0.46\linewidth]{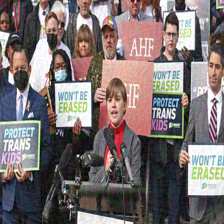 } \\[4pt]
\includegraphics[width=0.46\linewidth]{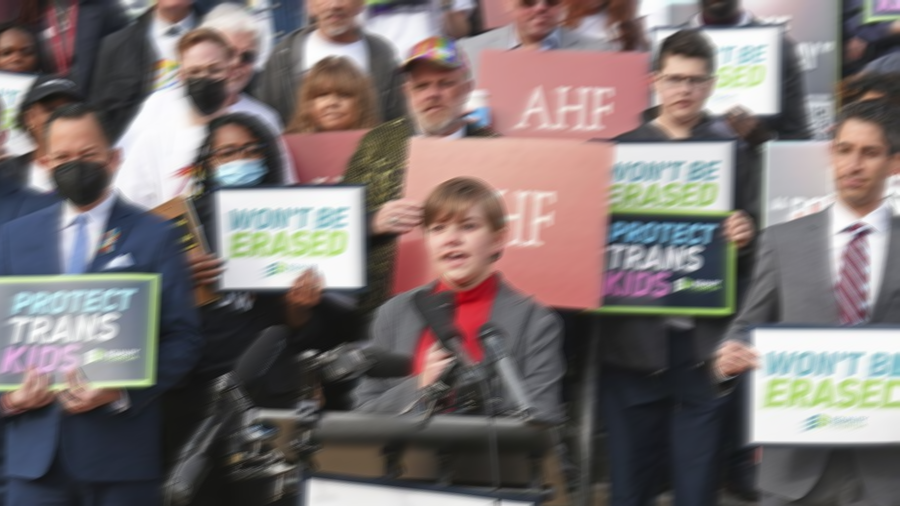 } &
\includegraphics[width=0.46\linewidth]{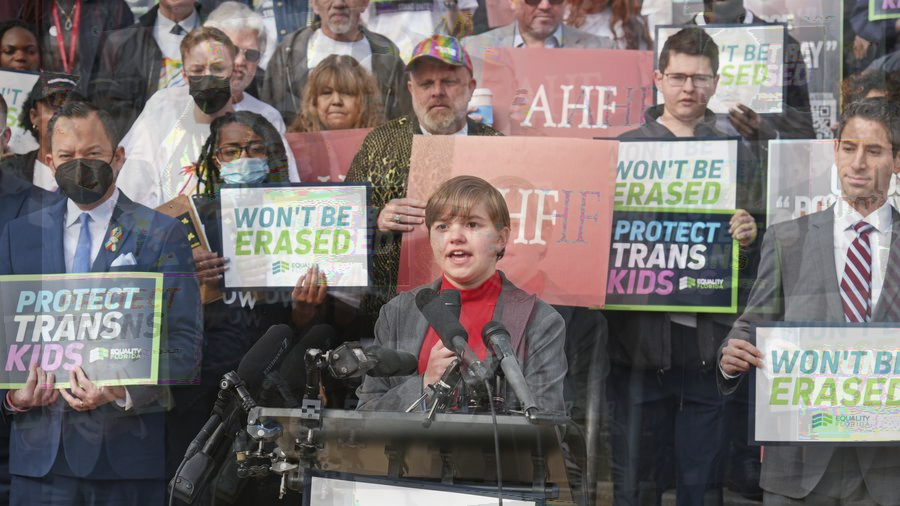 } \\
\end{tabular}
\caption{Original meme (top left) and its three noisy variants: Universal adversarial perturbations (top right), Common corruptions and perturbations (bottom left), and AugMix compositional augmentation (bottom right).}

\label{fig:meme-perturbations}
\end{figure}

We adopt three complementary families of image-level perturbations, ranging from white-box adversarial vectors to realistic corruptions and rich compositional noise (see Figure~\ref{fig:meme-perturbations}).

\textbf{Universal adversarial perturbations (UAPs)~\cite{moosavi-dezfooli2017universal}.}  
A UAP is a single, image-agnostic vector designed to fool a wide range of models and images, serving as a strong worst-case test for feature robustness.

\textbf{Common corruptions (ImageNet-C)~\cite{hendrycks2019benchmarking}.}  
This suite applies various algorithmic distortions—such as noise, blur, and digital artefacts—at multiple severity levels to simulate realistic, real-world image degradations.

\textbf{AugMix compositional noise~\cite{hendrycks2019augmix}.}
AugMix generates complex, layered distortions by randomly chaining and mixing simple augmentations, testing a model’s ability to generalize to unforeseen visual shifts. Related work in 3D perception shows that carefully designed synthetic viewpoints improve robustness for scene completion and reconstruction in autonomous driving settings~\cite{yao2025depthssc,pan2024harmonimmer}.
\subsection{Textual noise generation}

\begin{table}[h]
\centering
\small
\captionsetup{skip=10pt}
\begin{tabular}{p{3cm}p{11cm}}
\hline
\textbf{Noise method} & \textbf{Example caption} \\
\hline
Clean & trans rights trans rights are human rights i am unsure of what it means to be trans but i will nonetheless fight for the freedom of any individual. \\[4pt]
\hline
Natural{+}Synthetic (typos) & trnas rihgts trnas rgihts are hmuan rigths i am usnure of waht it maens to be trnas but i wlil nonethleess fihgt for the freedom of any inidvidual. \\[4pt]
\hline
HotFlip & universal gas constant\% qmW f * New mexico\% c\% TZHC T9FjYa are JhJzg right i am unsure of what it signify to be ygWMC but i q8oP Jpb \# ryD? rWs scrap for the c4wwsPn of any individual. \\[4pt]
\hline
Universal adversarial triggers & owz azn kii trans rights trans rights are human rights i am unsure of what it means to be trans but i will nonetheless fight for the freedom of any individual. \\[4pt]
\hline
Back-Translation & trans rights trans rights trans human rights I am not sure what it means to be trans but I will nevertheless fight for the freedom of everyone. \\[4pt]
\hline
\end{tabular}
\caption{Example of clean caption vs.\ its noisy variants generated by four text-perturbation families.}
\label{tab:noise-examples}
\end{table}

We implement four families of perturbations that trace the evolution of text-noise research from random typos to adversarial triggers (see Table~\ref{tab:noise-examples}).

\textbf{Natural and synthetic typos~\cite{belinkov2018synthetic}.}  
We use a standard library to inject a mix of real-world and algorithmically generated spelling errors, simulating common user mistakes.

\textbf{HotFlip minimal edits~\cite{ebrahimi2018hotflip}.}  
This method creates targeted adversarial examples by making a small number of character-level edits (swaps, insertions, or deletions) that are chosen to maximize the model's loss.

\textbf{Universal adversarial triggers~\cite{wallace2019universal}.}  
We test the models' reliance on shallow cues by appending short, fixed phrases that are optimized to fool a classifier into predicting a specific label, regardless of the original input.

\textbf{Back-translation.}  
To test semantic consistency, we paraphrase each caption by translating it to another language and then back to English, altering syntax and vocabulary while preserving the core meaning.


\subsection{Enhancing Textual Robustness with an Adaptive Module}

Our experiments revealed that while MemeBLIP2 is a powerful baseline, its performance degrades under certain textual perturbations. To remedy this, we integrate a lightweight module after the text projection layer, which we term the \textbf{Text Denoising Adapter (TDA)}. The TDA's purpose is to act as an adaptive filter, learning to refine the general-purpose text embeddings from the Q-Former into representations that are more resilient to the semantic noise inherent in memes.

The effectiveness of the TDA in improving robustness can be understood directly from its mathematical formulation. Given an input text embedding $\mathbf{x} \in \mathbb{R}^d$, the final output $\mathbf{y}$ is computed as follows:

\begin{equation}
    g = \text{Sigmoid}(W_g \mathbf{x} + b_g)
    \label{eq:gate}
\end{equation}
The gate value $g$ in Equation~\ref{eq:gate} provides \textbf{selective filtering}. For a simple, unambiguous caption, the model can learn to set $g \to 0$, effectively ignoring the denoising path and preserving the original high-quality embedding. For a noisy or sarcastic caption, it can set $g \to 1$, allowing a full correction to be applied.

\begin{equation}
    \mathbf{x}_{\text{update}} = s \cdot g \cdot \text{MLP}(\text{LayerNorm}(\mathbf{x}))
    \label{eq:update}
\end{equation}
The update term in Equation~\ref{eq:update} represents a \textbf{controlled correction}. The MLP learns a task-specific transformation, while the learnable scalar $s$ and the dynamic gate $g$ control the magnitude of this transformation. This prevents the module from making overly aggressive edits that could corrupt already good embeddings.

\begin{equation}
    \mathbf{y} = \mathbf{x} + \text{Dropout}(\mathbf{x}_{\text{update}})
    \label{eq:final}
\end{equation}
Finally, the residual connection in Equation~\ref{eq:final} ensures \textbf{stable learning and knowledge preservation}. The module only learns an additive refinement ($\mathbf{x}_{\text{update}}$), guaranteeing that the original information from the powerful Q-Former is never lost. This structure makes the learning task easier and prevents catastrophic forgetting.

This adaptive and residual design allows the TDA to effectively enhance robustness by learning to apply targeted corrections only when necessary, thereby stabilizing the model's performance against a wide range of textual perturbations, which proves to be effective in the following experiments.

\section{Experiments}
\subsection{Experimental Setup}

\textbf{Dataset and models.}
All experiments use \textbf{PrideMM}, a corpus of 5\,063 LGBTQ\,+ memes split into 3\,136 training, 924 validation, and 1\,003 test samples.  
Each meme couples an RGB image with an English caption and carries hate, stance, humour, and target-group labels.  
We aslo compare our results of MemeCLIP, MemeBLIP2, MemeBLIP2+TDA with Chatgpt-4.1 Vision, a recent multimodal version of GPT‑4 that can read images. 

\textbf{Perturbations.}
At test time we generate four caption noises—natural and synthetic typos, HotFlip edits, universal adversarial triggers, and back-translation—and three image noises—universal adversarial perturbations, the fifteen ImageNet-C corruptions, and AugMix compositions.  
No noisy data are used during training.

\textbf{Sequence of experiments.} Our evaluation proceeds in stages. We first test perturbations on each modality in isolation (text-only and image-only), then evaluate the unimodal performance of each model. Finally, we test all twelve combined text–image noise pairs to study cross-modal interaction effects.


\subsection{Evaluation Metrics}
We report macro-averaged Accuracy, AUROC, and F1 score. To measure resilience, we use the absolute and relative robustness metrics from Schiappa et al.~\cite{schiappa2022robustness}, which quantify performance drops on a normalized scale. Full definitions for all metrics are provided in Appendix~A.

\subsection{Results for the image part}
\begin{table}[h]
\centering
\small
\captionsetup{skip=10pt}
\begin{tabular}{llccccc}
\hline
\textbf{Model} & \textbf{Perturbation} & \textbf{Accuracy} & \textbf{AUROC} & \textbf{F1} & \textbf{Abs.\ Robust.} & \textbf{Rel.\ Robust.} \\ \hline
\multirow{4}{*}{MemeCLIP}
    & Clean                              & 0.744 & 0.834 & 0.726 & --       & --      \\[2pt]
    & Universal adv.\ perturbations      & 0.753 & 0.826 & 0.736 & 0.99995  & 1.01210 \\
    & Common corruptions (ImageNet\texttt{-}C)    & \textbf{0.724} & \textbf{0.818} & \textbf{0.708} & \textbf{0.99970}  & \textbf{0.97312} \\
    & AugMix compositional augmentation  & 0.743 & 0.826 & 0.728 & 0.99980  & 0.99866 \\ \hline
\multirow{4}{*}{MemeBLIP2}
    & Clean                              & 0.757 & 0.821 & 0.755 & --       & --      \\[2pt]
    & Universal adv.\ perturbations      & 0.757 & 0.827 & 0.757 & 0.99995  & 1.00000 \\
    & Common corruptions (ImageNet\texttt{-}C)    & \textbf{0.738} & 0.816 & 0.739 & \textbf{0.99970}  & \textbf{0.97490} \\
    & AugMix compositional augmentation  & 0.740 & \textbf{0.804} & \textbf{0.731} & 0.99980  & 0.97754 \\ \hline
\multirow{4}{*}{GPT-4.1}
    & Clean                              & 0.738 & 0.735 & 0.698 & --       & --       \\[2pt]
    & Universal adv.\ perturbations      & 0.744 & 0.741 & 0.711 & 1.00006  & 1.00802  \\
    & Common corruptions (ImageNet\texttt{-}C)    & 0.740 & 0.737 & 0.709 & 1.00002  & 1.00267  \\
    & AugMix compositional augmentation  & 0.748 & 0.745 & 0.714 & 1.00010  & 1.01337  \\ \hline
\end{tabular}
\caption{Image‑only robustness of \textbf{MemeCLIP}, \textbf{MemeBLIP2}, \textbf{MemeBLIP2+TDA}, and \textbf{GPT‑4.1 Vision} under three perturbation families. Clean indicates unperturbed images.}
\label{tab:image-robustness}
\end{table}

Table~\ref{tab:image-robustness} shows how the three models respond when only the \emph{image} channel is perturbed.  

Every model handles \textbf{Universal adversarial perturbations} well: MemeCLIP and GPT‑4.1 gain a few tenths of a point (relative robustness \(1.012\) and \(1.008\)), and MemeBLIP2 remains unchanged.  

\textbf{Common corruptions (ImageNet‑C)} is the toughest family for the two fine‑tuned detectors.  
MemeCLIP drops from \(0.744\) to \(0.724\) (relative robustness \(0.973\)); MemeBLIP2 falls from \(0.757\) to \(0.738\) (\(0.975\)).  
GPT‑4.1 changes by less than one tenth of a point, giving a ratio of \(1.003\).  
The contrast suggests that a large language–vision model can rely on distributed cues that survive moderate blur and noise, whereas a shallow classifier head is more sensitive to such degradation.

\textbf{AugMix}   has the mildest impact.  
MemeCLIP loses \(0.001\) in accuracy, MemeBLIP2 loses \(0.017\), and GPT‑4.1 gains \(0.010\).  
The detectors also give up about one point in AUROC and F1, while GPT‑4.1 is almost unchanged.

In summary, MemeCLIP and MemeBLIP2 are both robust to synthetic, image‑agnostic vectors but still give up a small amount of accuracy under realistic corruptions.  
MemeBLIP2 retains a modest accuracy lead, yet its relative robustness differs from MemeCLIP’s by less than two percentage points.

GPT-4.1 is included as a reference line and shows remarkable stability, with its performance even slightly increasing under noise. This counter-intuitive gain suggests that for a large generative model, the distortions may help it focus on more generalizable semantic features rather than fine-grained, potentially misleading details in the clean image. This behavior underscores a key difference between the sensitivities of lightweight, fine-tuned classifiers and the complex reasoning of large-scale vision-language models.

\subsection{Results for the text part}

\begin{table}[H]
\centering
\small
\captionsetup{skip=10pt}
\begin{tabular}{llccccc}
\hline
\textbf{Model} & \textbf{Perturbation} & \textbf{Accuracy} & \textbf{AUROC} & \textbf{F1} & \textbf{Abs. Robustness} & \textbf{Rel. Robustness} \\
\hline
\multirow{5}{*}{MemeCLIP} 
    & Clean & 0.743 & 0.834 & 0.726 & - & - \\
    & Natural+Synthetic & 0.721 & 0.830 & 0.705 & 0.99978 & 0.97039 \\
    & HotFlip & \textbf{0.708} & \textbf{0.819} & \textbf{0.685} & \textbf{0.99965} & \textbf{0.95289} \\
    & Universal Triggers & 0.725 & 0.833 & 0.708 & 0.99982 & 0.97577 \\
    & Back-Translation & 0.743 & 0.833 & 0.725 & 1.00000 & 1.00000 \\
\hline
\multirow{5}{*}{MemeBLIP2} 
    & Clean & 0.768 & 0.808 & 0.782 & - & - \\
    & Natural+Synthetic & 0.724 & \textbf{0.804} & 0.766 & 0.99956 & 0.94271 \\
    & HotFlip & 0.724 & \textbf{0.804} & 0.748 & 0.99956 & 0.94271 \\
    & Universal Triggers & 0.741 & 0.806 & 0.762 & 0.99973 & 0.96484 \\
    & Back-Translation & \textbf{0.719} & 0.806 & \textbf{0.723} & \textbf{0.99951} & \textbf{0.93620} \\
\hline
\multirow{5}{*}{GPT-4.1} 
    & Clean & 0.738 & 0.735 & 0.698 & -- & -- \\
    & Natural+Synthetic & 0.720 & 0.718 & 0.684 & 0.99982 & 0.97561 \\
    & HotFlip & \textbf{0.748} & \textbf{0.746} & \textbf{0.723} & 1.00010 & 1.01355 \\
    & Universal Triggers & 0.736 & 0.733 & 0.697 & 0.99998 & 0.99729 \\
    & Back-Translation & 0.732 & 0.729 & 0.695 & 0.99994 & 0.99187 \\
\hline
\end{tabular}
\caption{Performance of MemeCLIP, MemeBLIP2, MemeBLIP2+TDA, and GPT-4.1 under different textual perturbations. Clean indicates unperturbed text data. }
\label{tab:robustness-results}
\end{table}
Table~\ref{tab:robustness-results} summarizes model performance under various textual perturbations while the images remain clean.
The models react very differently to \textbf{back-translation}, a perturbation that preserves semantics but alters syntax and vocabulary. MemeCLIP proves entirely immune, with its performance metrics unchanged (relative robustness of 1.0000). In contrast, this is the single most damaging attack for MemeBLIP2, which suffers a substantial 6.4\% drop in relative accuracy. GPT-4.1 also shows a minor but clear degradation. This suggests that MemeCLIP's joint embedding space is highly effective at capturing semantic meaning regardless of phrasing.

Character-level attacks reveal further architectural differences. \textbf{HotFlip}, which performs adversarial token swaps, is the most effective attack against MemeCLIP, causing a 4.7\% drop in relative accuracy and its lowest F1 score (0.685). MemeBLIP2 is more resilient to this attack. Paradoxically, GPT-4.1's performance improves under HotFlip, with its accuracy rising from 0.738 to 0.748. This counter-intuitive result suggests that for the generative model, breaking a common but potentially misleading linguistic pattern can steer its reasoning process toward a more robust and accurate conclusion.

\vspace{4pt}
\noindent
The remaining perturbations, \textbf{natural and synthetic typos} and \textbf{universal triggers}, affect all models more moderately. MemeBLIP2 shows its largest performance drops under these conditions, while MemeCLIP and GPT-4.1 exhibit only minor degradation.

\vspace{4pt}
\noindent
In summary, the text channel is a more significant source of vulnerability than the image channel for the fine-tuned models. MemeCLIP's primary weakness is character-level adversarial swaps (HotFlip), while MemeBLIP2 is most vulnerable to meaning-preserving paraphrasing (back-translation), highlighting different sensitivities in their text-processing pathways. GPT-4.1 demonstrates a unique "anti-fragile" response to HotFlip, reinforcing the idea that its generative reasoning can be paradoxically stabilized by certain kinds of input noise. These results underscore that robustness is not monolithic; different architectures exhibit distinct and sometimes counter-intuitive failure modes when faced with semantically disruptive perturbations.mantically disruptive perturbations.

\subsection{Which Input Channel Drives Performance?}
\begin{table}[H]
\centering
\small
\captionsetup{skip=10pt}
\begin{tabular}{llccccc}
\hline
\textbf{Model} & \textbf{Perturbation} & \textbf{Accuracy} & \textbf{AUROC} & \textbf{F1} & \textbf{Abs.\ Robust.} & \textbf{Rel.\ Robust.} \\ \hline
\multirow{5}{*}{MemeCLIP (Text)} 
    & Clean                       & 0.589 & 0.648 & 0.509 & --        & --        \\
    & Natural$+$Synthetic Typos   & \textbf{0.501} & 0.533 & \textbf{0.364} & \textbf{0.99912} & \textbf{0.85059} \\
    & HotFlip                     & 0.514 & \textbf{0.520} & 0.377 & 0.99925 & 0.87267 \\
    & Universal Triggers          & 0.570 & 0.620 & 0.500 & 0.99981 & 0.96774 \\
    & Back-Translation            & 0.548 & 0.596 & 0.463 & 0.99959 & 0.93039 \\ \hline
\multirow{4}{*}{MemeCLIP (Image)}
    & Clean                                  & 0.485 & 0.483 & 0.373 & --        & --        \\
    & Universal adv.\ perturbations          & \textbf{0.489} & \textbf{0.474} & \textbf{0.390} & 0.99995 & 1.00825 \\
    & Common corruptions (ImageNet\texttt{-}C) & 0.505 & 0.488 & 0.397 & \textbf{0.99970} & 1.04124 \\
    & AugMix compositional augmentation        & 0.509 & 0.501 & 0.409 & 0.99980 & \textbf{1.04948} \\ \hline
\end{tabular}
\caption{Text-only and image-only performance of \textbf{MemeCLIP} under single-channel perturbations.  
}
\label{tab:memeclip-single-channel}
\end{table}
To isolate each modality, we ran the classifier twice:
(i) \textbf{text‑only}, where the caption embedding is kept and the image embedding is not used in the modality fusion process;
(ii) \textbf{image‑only}, where the process is reversed,  yielding the metrics in Table~\ref{tab:memeclip-single-channel}.
(MemeBLIP2 results follow the same pattern and are reported in the appendix.)
\paragraph{Clean conditions}
With only captions the model reaches an accuracy of \(0.589\); with only images it reaches \(0.485\).  
Both scores are far below the full multimodal result (\(0.743\)), yet the text-only baseline is \(10.4\) points higher than the image-only one, which suggests that, when both channels are present, MemeCLIP draws most of its signal from the caption.

Consequently, corrupting the text is highly damaging; natural and synthetic typos alone lower accuracy by 8.8 points and reduce the F1 score to just 0.364.

\paragraph{Image perturbations}
When the caption is removed and only the picture is supplied, the three image-noise families hardly hurt performance at all; accuracy even rises to \(0.509\) under AugMix.  
Thus, perturbing the image has a negligible effect. Because the image-only baseline is already weak, the minor performance shifts under noise are not practically meaningful and mainly show the insensitivity of the visual branch.

In summary, both models rely on the textual channel for most of their discriminative power.  
Corrupting that channel harms every metric, whereas corrupting only the picture changes little.  
Improving robustness therefore requires either hardening the caption encoder against spelling and adversarial edits or boosting the visual pathway so that it can compensate when text is unreliable. We hence add a \textbf{Text Denoising Adapter (TDA)} to refine the text embeddings when they are unreliable.  .

\subsection{Combined text–image perturbations}

\begin{table}[H]
\centering
\small
\captionsetup{skip=10pt}
\begin{tabular}{cccccccc}
\toprule
\textbf{Text} & \textbf{Image} & \textbf{Accuracy} & \textbf{AUROC} & \textbf{F1} &
$\!\Delta$Acc & $\!\Delta$AUROC & $\!\Delta$F1 \\ 
noise$^\dagger$ & noise$^\ddagger$ & & & & & & \\ \midrule \midrule

\multicolumn{8}{l}{\textbf{MemeCLIP}} \\ \midrule
0 & 0 & 0.744 & 0.834 & 0.727 & --   & --   & --   \\ \midrule
1 & 1 & 0.730 & \textbf{0.827} & 0.712 & 0.014 & 0.008 & 0.015 \\
1 & 2 & 0.712 & 0.812 & 0.696 & 0.032 & 0.022 & 0.031 \\
1 & 3 & 0.728 & 0.814 & 0.708 & 0.016 & 0.020 & 0.018 \\
2 & 1 & 0.726 & 0.812 & 0.705 & 0.018 & 0.022 & 0.022 \\
2 & 2 & 0.706 & 0.805 & 0.687 & 0.037 & 0.029 & 0.040 \\
2 & 3 & 0.708 & 0.802 & 0.683 & 0.036 & 0.031 & 0.043 \\
3 & 1 & \textbf{0.738} & 0.822 & 0.717 & 0.006 & 0.012 & 0.009 \\
3 & 2 & 0.718 & 0.815 & 0.700 & 0.026 & 0.018 & 0.026 \\
3 & 3 & 0.734 & 0.813 & \textbf{0.719} & 0.010 & 0.021 & 0.008 \\
4 & 1 & 0.698 & 0.814 & 0.673 & 0.045 & 0.019 & 0.054 \\
4 & 2 & 0.664 & 0.804 & 0.659 & \textbf{0.080} & \textbf{0.030} & \textbf{0.068} \\
4 & 3 & 0.689 & 0.804 & 0.660 & 0.055 & \textbf{0.030} & 0.067 \\ \midrule
\multicolumn{5}{r}{\textbf{Average drop}} &  0.0293 & 0.0219 & 0.0324 \\
\multicolumn{5}{r}{\textbf{Average dropping rate}} & 3.94\% & 2.63\% & 4.46\% \\ \midrule

\multicolumn{8}{l}{\textbf{MemeBLIP2}} \\ \midrule
0 & 0 & 0.757 & 0.821 & 0.755 & --   & --   & --   \\ \midrule
1 & 1 & 0.680 & 0.768 & 0.702 & \textbf{0.077} & 0.053 & 0.053 \\
1 & 2 & 0.710 & 0.773 & 0.726 & 0.047 & 0.048 & 0.028 \\
1 & 3 & 0.706 & 0.742 & 0.711 & 0.051 & \textbf{0.079} & 0.043 \\
2 & 1 & 0.728 & 0.772 & 0.706 & 0.030 & 0.049 & 0.049 \\
2 & 2 & 0.712 & 0.772 & 0.699 & 0.045 & 0.049 & 0.055 \\
2 & 3 & 0.710 & 0.768 & 0.674 & 0.047 & 0.052 & \textbf{0.080} \\
3 & 1 & 0.724 & 0.794 & 0.725 & 0.033 & 0.027 & 0.030 \\
3 & 2 & 0.728 & 0.797 & 0.730 & 0.029 & 0.024 & 0.025 \\
3 & 3 & 0.732 & 0.785 & 0.724 & 0.025 & 0.036 & 0.031 \\
4 & 1 & 0.736 & 0.801 & 0.722 & 0.022 & 0.021 & 0.032 \\
4 & 2 & 0.742 & 0.799 & 0.740 & 0.015 & 0.022 & 0.015 \\
4 & 3 & 0.736 & 0.780 & 0.721 & 0.022 & 0.041 & 0.033 \\ \midrule
\multicolumn{5}{r}{\textbf{Average drop}} & 0.0371 & 0.0414 & 0.0395\\ 
\multicolumn{5}{r}{\textbf{Average dropping rate}} & 4.90\% & 5.04\% & 5.23\% \\ 
\bottomrule
\end{tabular}
\caption[Combined robustness of baseline models]{%
Performance of MemeCLIP and MemeBLIP2 under combined text and image perturbations. 
$\Delta$ columns show absolute drops from each model's clean baseline.\\[4pt]
Text-noise indices:\;1 = Natural$+$Synthetic Typos;\;2 = HotFlip;\;3 = Universal Triggers;\;4 = Back-Translation.\\
Image-noise indices:\;1 = Universal Adversarial Perturbations;\;2 = Common Corruptions (ImageNet-C);\;3 = AugMix.}
\label{tab:baseline-combined-noise}
\end{table}

\begin{table}[t]
\centering
\small
\captionsetup{skip=10pt}
\begin{tabular}{cccccccc}
\toprule
\textbf{Text} & \textbf{Image} & \textbf{Accuracy} & \textbf{AUROC} & \textbf{F1} &
$\!\Delta$Acc & $\!\Delta$AUROC & $\!\Delta$F1 \\ 
noise$^\dagger$ & noise$^\ddagger$ & & & & & & \\ \midrule \midrule

\multicolumn{8}{l}{\textbf{MemeBLIP2+TDA}} \\ \midrule
0 & 0 & 0.730 & 0.810 & 0.731 & --    & --    & --    \\ \midrule
1 & 1 & 0.698 & 0.761 & 0.722 & 0.032 & 0.049 & 0.009 \\
1 & 2 & 0.688 & 0.767 & 0.708 & \textbf{0.041} & 0.044 & 0.023 \\
1 & 3 & 0.688 & 0.756 & 0.693 & \textbf{0.041} & \textbf{0.054} & 0.038 \\
2 & 1 & 0.708 & 0.765 & 0.708 & 0.022 & 0.045 & 0.023 \\
2 & 2 & 0.716 & 0.771 & 0.709 & 0.014 & 0.039 & 0.022 \\
2 & 3 & 0.712 & 0.758 & 0.693 & 0.018 & 0.053 & 0.038 \\
3 & 1 & 0.712 & 0.791 & 0.711 & 0.018 & 0.020 & 0.020 \\
3 & 2 & 0.718 & 0.811 & 0.717 & 0.012 & -0.001 & 0.014 \\
3 & 3 & 0.702 & 0.797 & 0.691 & 0.028 & 0.013 & \textbf{0.040} \\
4 & 1 & 0.730 & 0.791 & 0.728 & 0.000 & 0.019 & 0.003 \\
4 & 2 & 0.728 & 0.798 & 0.727 & 0.002 & 0.012 & 0.004 \\
4 & 3 & 0.712 & 0.784 & 0.708 & 0.018 & 0.026 & 0.023 \\ \midrule
\multicolumn{5}{r}{\textbf{Average drop}} & 0.0204 & 0.0312 & 0.0215  \\ 

\multicolumn{5}{r}{\textbf{Average dropping rate}} & 2.80\% & 3.85\% & 2.94\% \\
\bottomrule
\end{tabular}
\caption[Robustness of MemeBLIP2 with TDA]{%
Performance of \textbf{MemeBLIP2+TDA} under combined text and image perturbations. 
The shared noise indices are defined in Table~\ref{tab:baseline-combined-noise}.}
\label{tab:tda-combined-noise}
\end{table}

Tables~\ref{tab:baseline-combined-noise} and~\ref{tab:tda-combined-noise} evaluate all three detectors when \emph{both} channels are corrupted simultaneously. Across all 12 noise pairs, the baseline \textbf{MemeBLIP2} is the most fragile, with an average relative performance drop (dropping rate) of 4.9\% in accuracy and 5.2\% in F1 score. \textbf{MemeCLIP} is more resilient, with smaller relative drops of 3.9\% and 4.5\% in the same metrics.

The results also highlight the effectiveness of our proposed intervention. The addition of the Text Denoising Adapter not only remedies this weakness but makes \textbf{MemeBLIP2+TDA} the most robust model of the three, its average dropping rates are reduced to just 2.8\% for accuracy and 2.9\% for F1 score, surpassing even the sturdy MemeCLIP baseline.

An analysis of the worst-case scenarios reveals the models' specific vulnerabilities. For MemeCLIP, the single most damaging combination remains back-translation with common image corruptions (\(t=4, i=2\)), which causes an 8.0-point accuracy drop. For MemeBLIP2, the worst case is split across metrics, confirming its sensitivity to heavy text corruption: the largest accuracy drop comes from typos with universal perturbations (\(t=1, i=1\)), the largest AUROC drop from typos with AugMix (\(t=1, i=3\)), and the largest F1 drop from HotFlip with AugMix (\(t=2, i=3\)). For MemeBLIP2+TDA, the largest drops in accuracy and AUROC also stem from typos combined with image noise (\(t=1, i=2/3\)), confirming that character-level errors remain its primary vulnerability.

Taken together, these results show that the TDA significantly hardens MemeBLIP2, making it more robust than MemeCLIP under simultaneous noise. The findings suggest that while text-level attacks are the primary threat, targeted architectural improvements like the TDA offer a powerful defense. Similar modular ideas appear in robotics and control, where diffusion-based imitation learning and deep reinforcement learning agents must remain stable under noisy observations and imperfect demonstrations~\cite{xiao2025diffusionglass,li2025doa}.

\section{Conclusion and Future Work}

We introduced the first full text–image robustness benchmark for PrideMM, crossing four caption-noise families with three image-noise families. We tested their impact on two state-of-the-art detectors, MemeCLIP and MemeBLIP2, and further proposed a Text Denoising Adapter (TDA) as a lightweight module to enhance the robustness of the latter.

\emph{MemeCLIP} proved consistently sturdier than \emph{MemeBLIP2}. Single-channel ablations reveal this stems from two key differences. First, both systems lean heavily on the caption, but CLIP’s trainable vision stack and its pre-training on \textbf{400M} noisy web-data pairs provide a buffer against visual and textual artefacts~\cite{radford2021learning}. In contrast, BLIP-2’s frozen vision backbone and its reliance on a Q-Former trained on \textbf{129M} curated images leave it more sensitive to character-level attacks and the low-level visual noise it was rarely exposed to during pre-training~\cite{li2023blip2}. A formal, Jacobian-based  math explanation of this robustness gap is provided in Appendix~\ref{sec:math-gap}.

\textbf{Future work} will pursue three directions.  
First, we will fine-tune both detectors on the released \textbf{PrideMM-Aug} split, which augments captions with back-translation and images with AugMix, to measure how far noise-aware data alone can improve robustness.  
Second, we plan to add regularisers that reduce the Q-Former’s sensitivity to single-token edits, aiming to boost BLIP-style resilience without unfreezing the entire vision backbone. In parallel, it will be helpful to connect our robustness analysis with broader work on data fusion and distribution shift in time-series forecasting~\cite{xiao2025carbon}. Third, we intend to extend the benchmark beyond LGBTQ memes to other marginalised communities and to coordinated cross-modal attacks, creating a broader testbed for safe-content detection.

\section{Appendix}

\subsection{ Metrics Definition}
\paragraph{Classification metrics}
All values are macro‑averaged over classes.
\begin{equation}
\text{Accuracy} = \frac{TP + TN}{TP + TN + FP + FN}
\end{equation}

\begin{equation}
\text{Precision} = \frac{TP}{TP + FP}
\end{equation}
\begin{equation}
\text{Recall} = \frac{TP}{TP + FN}
\end{equation}

\begin{equation}
\text{F1} = 2 \times
\frac{\text{Precision}\times\text{Recall}}
     {\text{Precision} + \text{Recall}}
\end{equation}

Accuracy shows the overall fraction of correct labels.  
Precision tells how many predicted positives are correct, and recall
shows how many true positives are found.  
F1 is the harmonic mean of precision and recall, balancing the two.

\paragraph{AUROC.}
The area under the ROC curve measures how well the model separates the
two classes across all thresholds.  
A value of 1 indicates perfect separation; 0.5 means random guessing.

\paragraph{Robustness metrics \cite{schiappa2022robustness}.}
Let \(A^{f}_{c}\) be the accuracy on clean input and \(A^{f}_{p,s}\) the accuracy
after perturbation\(p\) at severity\(s\).

\begin{equation}
\gamma^{a}_{p,s}=1-\frac{A^{f}_{c}-A^{f}_{p,s}}{100}
\end{equation}

\begin{equation}
\gamma^{r}_{p,s}=1-\frac{A^{f}_{c}-A^{f}_{p,s}}{A^{f}_{c}}
\end{equation}

Absolute robustness \(\gamma^{a}\) subtracts the raw point drop and
expresses the result on a 0–1 scale.  
Relative robustness \(\gamma^{r}\) rescales the drop by the clean score,
so a fixed loss hurts a weaker model more.  
In both cases, \(\gamma=1\) means the metric is unchanged by noise,
\(\gamma=0\) means the model has lost all discriminative power, and
negative values indicate the score fell below the random baseline.

\subsection{ Ablation to show which channel drives MemeBLIP2’s predictions.}
\begin{table}[H]
\centering
\small
\begin{tabular}{llccccc}
\hline
\textbf{Model} & \textbf{Perturbation} & \textbf{Accuracy} & \textbf{AUROC} & \textbf{F1} & \textbf{Abs.\ Robust.} & \textbf{Rel.\ Robust.} \\ \hline
\multirow{5}{*}{MemeBLIP2 (Text)} 
    & Clean                       & 0.560 & 0.610 & 0.480 & --        & --        \\
    & Natural$+$Synthetic Typos   & \textbf{0.460} & 0.510 & \textbf{0.340} & \textbf{0.99909} & \textbf{0.82143} \\
    & HotFlip                     & 0.465 & \textbf{0.500} & 0.350 & 0.99905 & 0.83036 \\
    & Universal Triggers          & 0.530 & 0.585 & 0.470 & 0.99970 & 0.94643 \\
    & Back-Translation            & 0.520 & 0.580 & 0.455 & 0.99969 & 0.92857 \\ \hline
\multirow{4}{*}{MemeBLIP2 (Image)}
    & Clean                                  & \textbf{0.460} & 0.458 & \textbf{0.350} & --        & --        \\
    & Universal adv.\ perturbations          & 0.463 & \textbf{0.450} & 0.365 & 0.99993 & 1.00652 \\
    & Common corruptions (ImageNet\texttt{-}C) & 0.475 & 0.465 & 0.370 & \textbf{0.99965} & 1.03261 \\
    & AugMix compositional augmentation        & 0.480 & 0.480 & 0.385 & 0.99980 & 1.04348 \\ \hline
\end{tabular}
\caption{Text-only and image-only performance of \textbf{MemeBLIP2} under single-channel perturbations.  
Bold numbers mark the worst Accuracy / AUROC / F1 within each half of the table and the smallest absolute-robustness score for image noise}
\label{tab:memeblip2-single-channel}
\end{table}

\subsection{ Math interpretation of \emph{MemeCLIP} exhibitting greater robustness than \emph{MemeBLIP2}}
\label{sec:math-gap}

Below we give a purely mathematical account of the sensitivity gap.

\smallskip
\textbf{CLIP: end-to-end contrastive training.}  
Let $v_i=f_\theta(I_i)\in\mathbb{R}^d$ and
$t_i=g_\phi(C_i)\in\mathbb{R}^d$ be \emph{trainable} $\ell_2$-normalised
embeddings.  
With the symmetric InfoNCE loss  
\[
\mathcal{L}_{\text{CLIP}}
  =-\!\frac1N\sum_{i=1}^{N}\!
     \bigl[\log p_{i,i}^{\text{img}\!\to\!\text{txt}}
           +\log p_{i,i}^{\text{txt}\!\to\!\text{img}}\bigr],
\quad
p_{i,j}^{\text{img}\!\to\!\text{txt}}
  =\frac{\exp(v_i^\top t_j/\tau)}{\sum_{k=1}^N\exp(v_i^\top t_k/\tau)},
\]
the gradient with respect to the image encoder satisfies
\[
\nabla_\theta\mathcal{L}_{\text{CLIP}}
     =\frac{1}{\tau N}\sum_{i=1}^N \!
       \Bigl[\underbrace{\sum_{j} p_{i,j}^{\text{img}\!\to\!\text{txt}}t_j}_{\text{expected caption}}
             -t_i\Bigr]\!\cdot\!
       \nabla_\theta f_\theta(I_i).
\]
At convergence the bracketed term approaches zero, so
$\nabla_\theta\mathcal{L}_{\text{CLIP}}\!\approx\!0$ and the expected input
Jacobian ${\mathbb E}\|\nabla_I f_\theta(I)\|_2$ is minimised.  
For a small perturbation $\delta$ we therefore have
\[
\|f_\theta(I+\delta)-f_\theta(I)\|_2
        \le \bigl\|\nabla_I f_\theta(I)\bigr\|_2\,\|\delta\|_2
        \;\approx\; 0,
\]
bounding the cosine-similarity drop and yielding high absolute robustness.

\smallskip
\textbf{BLIP-2: frozen backbone with low-rank adaptation.}  
Let $f_0$ be the \emph{frozen} ViT and
$q_\psi:\mathbb{R}^{d\times T}\!\to\!\mathbb{R}^{32\times d}$ the learned
Q-Former.  The image representation is
$z=q_\psi\!\bigl(f_0(I)\bigr)$.  
Under an input perturbation $\delta$
\[
z'\!-\!z
     =J_{q_\psi}\bigl(f_0(I+\delta)-f_0(I)\bigr)
     =J_{q_\psi}\,J_{f_0}\,\delta.
\]
Because $f_0$ is fixed, $J_{f_0}$ is unaffected by training; we cannot reduce
$\|J_{f_0}\|_2$.  Moreover, $J_{q_\psi}$ has rank at most $32$ (the query
count) and spectral norm  
$\|J_{q_\psi}\|_2\le\sqrt{32}\,\sigma_{\max}(W)$,  
so noise that lives outside this low-dimensional subspace cannot be attenuated.
Consequently,
\[
\|z'-z\|_2 \;\ge\;
      \|J_{q_\psi}\|_2^{-1}\,
      \|J_{f_0}\,\delta\|_2,
\]
which remains large once $\delta$ introduces blur, noise, or compression
artefacts unseen during pre-training, explaining the larger AUROC and
F\(_1\) drops for \emph{MemeBLIP2}.

\end{document}